\documentclass[submitting]{nst}

\usepackage{dcolumn}
\usepackage{mwe}
\usepackage{multirow}
\usepackage{braket}
\usepackage{color}
\usepackage{float}
\usepackage[T2A,T1]{fontenc}
\usepackage[russian,english]{babel}
\usepackage{verbatim}

\usepackage{listings}
\preto{\nolinenumbers}

\begin{document}  

\title{Prediction of Nuclear Charge Density Distribution with Feedback Neural Network}

\author{Tian-Shuai Shang}
\affiliation{College of Physics, Jilin University, Changchun 130012, China}
\author{Jian Li}
\email{E-mail: jianli@jlu.edu.cn}
\affiliation{College of Physics, Jilin University, Changchun 130012, China}
\author{Zhong-Ming Niu}
\email{E-mail: zmniu@ahu.edu.cn}
\affiliation{School of Physics and Optoelectronic Engineering, Anhui University, Hefei 230601, China}

\begin{abstract}  
  The nuclear charge density distribution plays an important role in nuclear physics and atomic physics. As one of the most frequently used models to obtain charge density distribution, the two-parameter fermi (2pF) model has been widely applied in both nuclear physics and atomic physics. Currently, the feedforward neural network has been employed to study the available 2pF model parameters for 86 nuclei, and it is found that by introducing A1/3 into the input parameter of the neural network, the accuracy and precision of the parameter learning effect are improved. Furthermore, the average result of multiple predictions is more reliable than the best result of a single prediction, and there is no significant difference between the average result of density value and of parameter value for the average charge density distribution. In addition, 2pF parameters of 284 (near) stable nuclei are also predicted in this work, which provides a reference for the experiment.
\end{abstract}

\keywords{charge density distribution, two-parameter Fermi model, feedforward neural network approach}

\maketitle

\section{Introduction}\label{sec1}
Ever since Rutherford discovered the nucleus \cite{RutherfordE.:1911}, there has been an interest in studying the charge density distribution of the nucleus, 
because density distribution information is very important to analyze the nuclear structure. 
In fact, the nuclear charge density distributnion gives direct information on the coulomb energy of nucleus, 
which allows one to use it to calculate the charge radius. Additionally, in studies of high-momentum tails (HMT) caused by 
short-range correlations (SRC), the percentage of tails is closely related to the distribution of nuclear charge density \cite{YangZ.X.andShangX.L.andYongG.C.andZuoW.andGaoY.:2019,ShangX.L.andDongJ.M.andZuoW.andYinP.andLombardoU.:2021}. 
Accurate estimation of neutron and proton density distributions is crucial in the study of asymmetric nuclear matter in nuclear
astrophysics, and the nuclear symmetry energy and its density dependence play an important role in understanding the physics 
of many terrestrial nuclear experiments and astrophysical observations \cite{HorowitzC.J.:2019,ChenY.J.:2019}. From the viewpoint of configurational information entropy (CIE), the charge distributions in the projectile fragmentation reaction 
may be good probes for determining theneutron-skin thickness of neutron-rich nuclei \cite{NST_Ma:2022}, and others like density dependence of symmetry energy \cite{NST_LiBing:2022,NST_LiLi:2022} and so on. 
On the other hand, the charge density  distribution 
also has a high status in atomic physics \cite{Andrae2017,PatoaryA.M.andOreshkinaN.S.:2018}. For example, if the charge density distribution in the nucleus is known, so we can 
calculate the deformation of the nucleus, and then get its influence on the electrons in the atom \cite{VISSCHER1997207,ANDRAE2000413}. 

Hofstadter measured the charge density of protons in the 1950s and described the density distribution of some nuclei 
on that basis \cite{HofstadterR.:1956}. So far, electron scattering experiments (elastic and inelastic) has become an effective method to measure the nuclear structure 
\cite{Ehrenberg1959,Kim:1992,Meyer-Berkhout1959,NST_WeiNan:2021}.
After obtaining density distribution data, two methods can be used to describe the shape of charge density, namely 
model-dependent analysis (e.g., two/three parametric Fermi model and two/three parametric Gaussian model) and model-independent 
analysis (e.g., Fourier Bessel analysis and Gaussian sum analysis) \cite{ChuY.Y.2011}. However, there are still few experimental data on 
nuclear charge density. Taking the model-related analysis method as an example, there are only less than 300 nuclei with charge density parameters confirmed 
\cite{DeJagerC.W.andDeVriesH.andDeVriesC.:1974,DeVriesH.andDeJagerC.W.andDeVriesC.:1987,FrickeG.andBernhardtC.andHeiligK.andSchallerL.A.andSchellenbergL.andSheraE.B.andDejagerC.W.:1995}, which are mainly concentrated near stable nuclei.

In recent decades, many microscopic models of nuclear structure have been successfully established, and almost all of them can 
calculate density distribution information, such as the $ab-initio$ (Green's function Monte Carlo method \cite{Carlson2015QuantumMC}, the self-consistent Green's function method \cite{Dickhoff_2004}, the coupled-cluster method \cite{Hagen_2014}, the lattice chiral effective field theory \cite{LEE2009117}, and the nocore shell model) and density functional theory (DFT) \cite{RevModPhys.75.121}. Both of them can accurately describe nuclear ground-state properties. 
However, as nuclear mass number increases, the expansion 
of configuration space limits the calculation range of $ab-initio$ and shell model. The systematicity of calculation 
and how to accurately describe it have become great challenge. The density functional theory, such as Skyrme-Hartree-Fock (SHF) 
method \cite{RichterW.A.:2003,AbbasS.A.:2022,AbdullahA.N.:2017} and Covariant Density Functional Theory (CDFT) \cite{RING1996193,VRETENAR2005101,MENG2006470,LiJ.:2018,MengJ.andPengJ.andZhangS.-Q.andZhaoP.-W.:2013,SHEN2019103713,doi:10.1142/9872}, are the most widely used and effective microscopic models for studying nuclear properties. The DFT with a small 
number of parameters allows a very successful description of ground-state properties of nuclei all over 
the nuclear chart \cite{RevModPhys.75.121}. Although its calculation range on nuclear chart is far beyond that of $ab-initio$ and shell models, 
its prediction of charge density distribution is not accurate enough sometimes.

Compared with the above microscopic theoretical models, empirical models are more commonly used to describe the distribution of nuclear 
charge density, such as the Fermi model and the Gaussian model in the model dependent analysis method, among which the 
two-parameter Fermi model (2pF) is one of the most widely used. 2pF can describe 
the stability of the central density of larger nuclei, and show the exponential decay of the surface density. More importantly, 2pF model only 
needs two parameters to describe the nuclear charge density and is easy to use. However, only a few nuclei have 2pF parameter 
according to Refs. \cite{DeJagerC.W.andDeVriesH.andDeVriesC.:1974,DeVriesH.andDeJagerC.W.andDeVriesC.:1987,FrickeG.andBernhardtC.andHeiligK.andSchallerL.A.andSchellenbergL.andSheraE.B.andDejagerC.W.:1995}, and other nuclei need to be extrapolated from existing experimental data. This can 
be done easily using machine learning methods.

Machine Learning (ML) is a science that studies how to use computers to simulate or realize human learning activities. It is one of the most 
intelligent and cutting-edge research fields in artificial intelligence (AI). In the past decades, the prodigious development of  machine  
learning  applications  has   impacted  many  fields  such  as  image  recognition \cite{HeK.M.andZhangX.Y.andRenS.Q.andSunJ.andIeee2016, HuangG.andLiuZ.andvanderMaatenL.andWeinbergerK.Q.andIeee2017} and  language  
translation \cite{DzmitryB.andKyunghyunC.andYoshuaB.:2015, BaroniM.andBernardiniS.:2005}. There are many algorithms that have been invented to make machine learning as close to the human mind 
as possible, the core of which is  the Back Propagation (BP) algorithm, which is the most powerful and popular machine learning 
tool. Other algorithms, such as Decision Tree (DT), Naive Bayesian Model (NBM), SVM and Cluster, have also been used in many areas,
providing powerful tools for particle physics \cite{BaldiP.andSadowskiP.andWhitesonD.:2014,PangL.G.andZhouK.andSuN.andPetersenH.andStockerH.andWangX.N.:2018,BrehmerJ.andCranmerK.andLouppeG.andPavezJ.:2018}, condensed matter physics \cite{CarrasquillaJ.andMelkoR.G.:2017,CarleoG.andTroyerM.:2017} and so on.

In the 1990s, ML with neural networks began to be applied to the modeling of observation data in nuclear physics 
\cite{GAZULA19921,GERNOTH19931} and have been widely used in various fields, such as ground state properties of nuclei 
including nuclear mass (binding energy), stability, separation energy, branching ratio of radioactive decay, etc 
\cite{NIU201848,PhysRevC.106.L021303,GAZULA19921,GERNOTH19931,Clark1999HigherorderPP,ATHANASSOPOULOS2004222,doi:10.1142/S0217979206036053,GERNOTH1995291,COSTIRIS2009,NST_MingXC:2022,NST_GaoZP:2021}. 
Others such as excited state \cite{WangY.F.NiuZ.M.ZhangX.Y.:2022,AkkoyunS.andKayaH.andTorunY.:2022,
LasseriR.-D.andRegnierD.andEbranJ.-P.andPenonA.:2020,
AkkoyunS.andLaouetN.andBenrachiF.:2020}, charge radius \cite{UtamaR.andChenW.-C.andPiekarewiczJ.:2016,
MaY.andSuC.andLiuJ.andRenZ.andXuC.andGaoY.:2020,WuD.andBaiC.L.andSagawaH.andZhangH.Q.:2020}, $\alpha$ decay 
\cite{RodriguezU.B.andVargasC.Z.andGoncalvesM.andDuarteS.B.andGuzmanF.:2019,
BanosRodriguezU.andZunigaVargasC.andGoncalvesM.andBarbosaDuarteS.andGuzmanF.:2019}, 
$\beta$ decay \cite{NiuZ.M.andLiangH.Z.andSunB.H.andLongW.H.andNiuY.F.:2019,CostirisN.J.andMavrommatisE.andGernothK.A.andClarkJ.W.:2009,
CostirisN.J.andMavrommatisE.andGernothK.A.andClarkJ.W.andLiH.:2008}, magnetic moment \cite{YuanZ.andTianD.andLiJ.andNiuZ.:2021}, nuclear reactions and cross-sections \cite{osti_1800793,2020ChPhC..44l4107M,Peng_2022,MaC.-W.andWeiH.-L.andLiuX.-Q.andSuJ.andZhengH.andLinW.-P.andZhangY.-X.:2021,Ma_2022}
nuclear structure data \cite{Akkoyun_2013,BAYRAM2014172,doi:10.1142/S0218301314500645,UtamaR.andPiekarewiczJ.andProsperH.B.:2016,NeufcourtL.andCaoY.andNazarewiczW.andViensF.:2018}, 
giant dipole resonance \cite{PhysRevC.104.034317}, $\beta$ decay one-neutron emission probabilities \cite{WUDI2021}, density functionals for nuclear systems \cite{PhysRevC.105.L031303}, nuclear data evaluation \cite{NST_Alhassan:2022} and so on. 
Among them, feedforward neural network (FNN) is widely used. \cite{GAZULA19921,GERNOTH19931,Clark1999HigherorderPP,ATHANASSOPOULOS2004222,
doi:10.1142/S0217979206036053,GERNOTH1995291,COSTIRIS2009,WuD.andBaiC.L.andSagawaH.andZhangH.Q.:2020,
LasseriR.-D.andRegnierD.andEbranJ.-P.andPenonA.:2020,AkkoyunS.andLaouetN.andBenrachiF.:2020,
CostirisN.J.andMavrommatisE.andGernothK.A.andClarkJ.W.andLiH.:2008}. It shows great learning ability during research because 
it can learn any function by adjusting the appropriate hyperparameters. Therefore, FNN will be adopted to study the 2pF 
experimental data and make predictions for other nuclei.

In this study, the basic formulas of the FNN approach are given in Sec. \ref{sec2}, the prediction results of charge density distribution 
are discussed in Sec. \ref{sec3}, and the summary and perspectives are presented in Sec. \ref{sec4}.

\section{TWO-PARAMETER FERMI MODEL AND FEEDFORWARD NEURAL NETWORK APPROACH} \label{sec2}
\subsection{TWO-PARAMETER FERMI MODEL}

As for the most cases, two-parameter Fermi distribution
\begin{equation}\label{eq1}
  \rho_{2pF} = \frac{N_0}{1 + e^{\frac{r - c}{z}}}
\end{equation}
is assumed for the charge distribution. The parameter $c$ is the half-density radius and $z$ is the diffuseness of the nuclear surface. $N_0$ 
is the normalization factor to satisfy
\begin{equation}\label{eq2}
  Z = \int_0^{\infty}\rho_{2pF}4 \pi r^2 dr,
\end{equation}
where $Z$ is the number of protons.

\subsection{FEEDFORWARD NEURAL NETWORK APPROACH}
\begin{figure}[H]  
  \includegraphics
  [width=1.0\hsize]
  {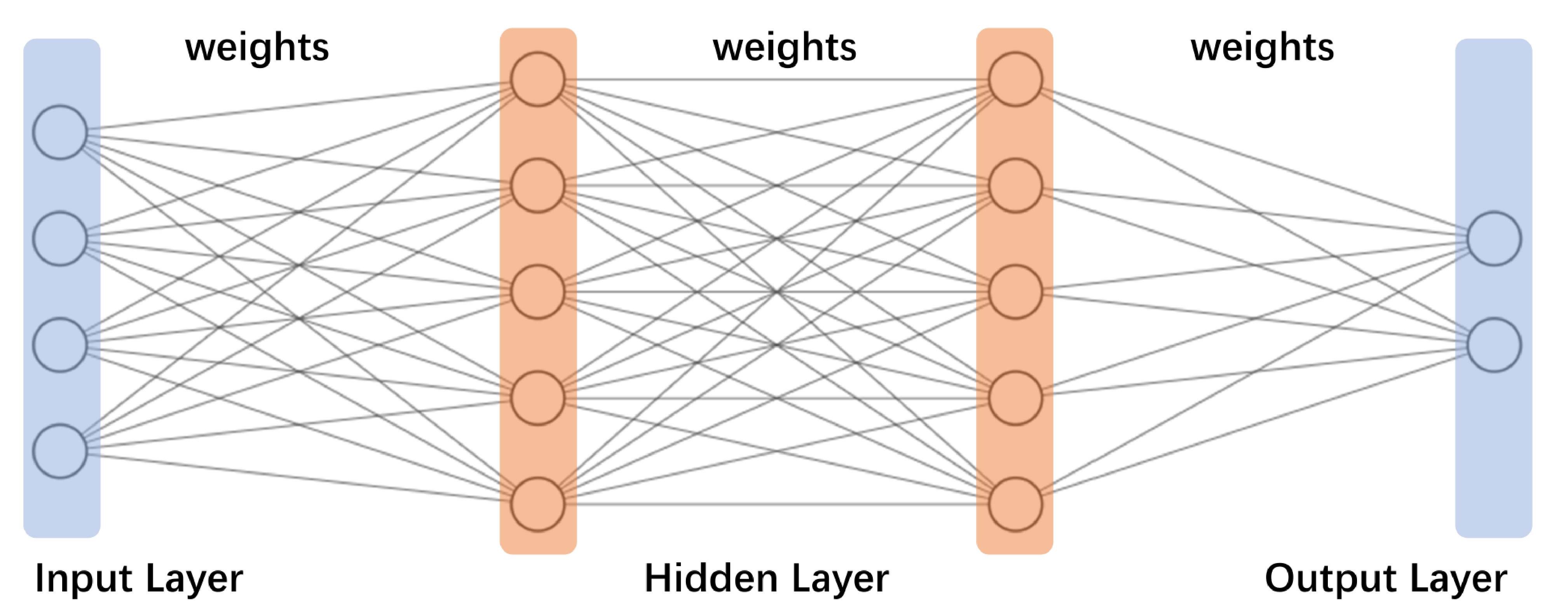}
  \caption{(Color online) A schematic diagram for a neural network with four input variables, two output variables, and two hidden layers (five neurons $H=5$ in each hidden layer).
  There also have four input variables and two output variables.}
  \label{fig:Figure1}
\end{figure}

Feedback neural networks is one of ML sub area. FNN mimics the human brain functionality in order to give 
outputs as consequence of the computation of the inputs. It is composed of processing units called neurons which have adaptive 
synaptic weights \cite{AkkoyunS.andLaouetN.andBenrachiF.:2020}. The framework of FNN is shown in Fig. \ref{fig:Figure1}, which is a multilayer neural network consisting 
of an input layer, hidden layers, and output layer. The number of hidden layers can vary, and neurons are fully connected between 
layers.

\begin{table*}[!htb]  
  \caption{MSE statistics of parameter $c$ and  $z$ on training set and validation set, and the datasets are randomly divided.
  SD represents the standard deviation calculated from Eq. (\ref{eq6}). Count is the number of results whose error is within the acceptable range (0.2 for $c$ and 0.005 for $z$)}
  \label{tab:Table1}
  \begin{tabular*}{12cm}{@{\extracolsep{\fill}} lcccccr}
    \toprule
    \specialrule{0em}{1pt}{1pt}
    \textbf{parameter c} & input                    & mean ($\si{\femto\metre}^2$)    & SD ($\si{\femto\metre}^2$)     & Count \\\specialrule{0em}{1pt}{1pt}
    \midrule
    \multirow{2}{*}{training set} & $Z, N, Z^{1/3}$          & 0.039 & 0.037 & \multirow{2}{*}{899}  \\
                         & $Z, N, Z^{1/3}, A^{1/3}$ & 0.024 & 0.031 &      \\
    \midrule
    \multirow{2}{*}{validation set}& $Z, N, Z^{1/3}$          & 0.037 & 0.035 & \multirow{2}{*}{891}  \\
                         & $Z, N, Z^{1/3}, A^{1/3}$ & 0.023 & 0.028 &      \\
    \toprule
    \toprule
    \specialrule{0em}{1pt}{1pt}
    \textbf{parameter z} & input                    & mean ($\si{\femto\metre}^2$)     & SD ($\si{\femto\metre}^2$)      & Count \\\specialrule{0em}{1pt}{1pt}
    \midrule
    \multirow{2}{*}{training set} & $Z, N, Z^{1/3}$          & 0.001 & 0.00030 & \multirow{2}{*}{998}  \\
                         & $Z, N, Z^{1/3}, A^{1/3}$ & 0.001 & 0.00027 &      \\
    \midrule
    \multirow{2}{*}{validation set}& $Z, N, Z^{1/3}$          & 0.002 & 0.00027 & \multirow{2}{*}{998}  \\
                         & $Z, N, Z^{1/3}, A^{1/3}$ & 0.002 & 0.00024 &      \\
    \bottomrule
  \end{tabular*}
\end{table*}
\begin{table*}[!htb]  
  \caption{Mean squared error statistics of parameter $c$ on training set and validation set, and the dataset is fixed.}
  \label{tab:Table2}
  \begin{tabular*}{12cm}{@{\extracolsep{\fill}} lcccccr} 
    \toprule
    \specialrule{0em}{1pt}{1pt}
    \textbf{parameter c} & input                    & mean ($\si{\femto\metre}^2$)    & SD ($\si{\femto\metre}^2$)     & Count \\\specialrule{0em}{1pt}{1pt}
    \midrule
    \multirow{2}{*}{training set} & $Z, N, Z^{1/3}$          & 0.035 & 0.031 & \multirow{2}{*}{894}  \\
                         & $Z, N, Z^{1/3}, A^{1/3}$ & 0.024 & 0.030 &      \\
    \midrule
    \multirow{2}{*}{validation set}& $Z, N, Z^{1/3}$          & 0.038 & 0.033 & \multirow{2}{*}{900}  \\
                         & $Z, N, Z^{1/3}, A^{1/3}$ & 0.028 & 0.032 &      \\
    \bottomrule
  \end{tabular*}
\end{table*}
\begin{figure*}[!htb]  
  \includegraphics
  [width=0.9\hsize]
  {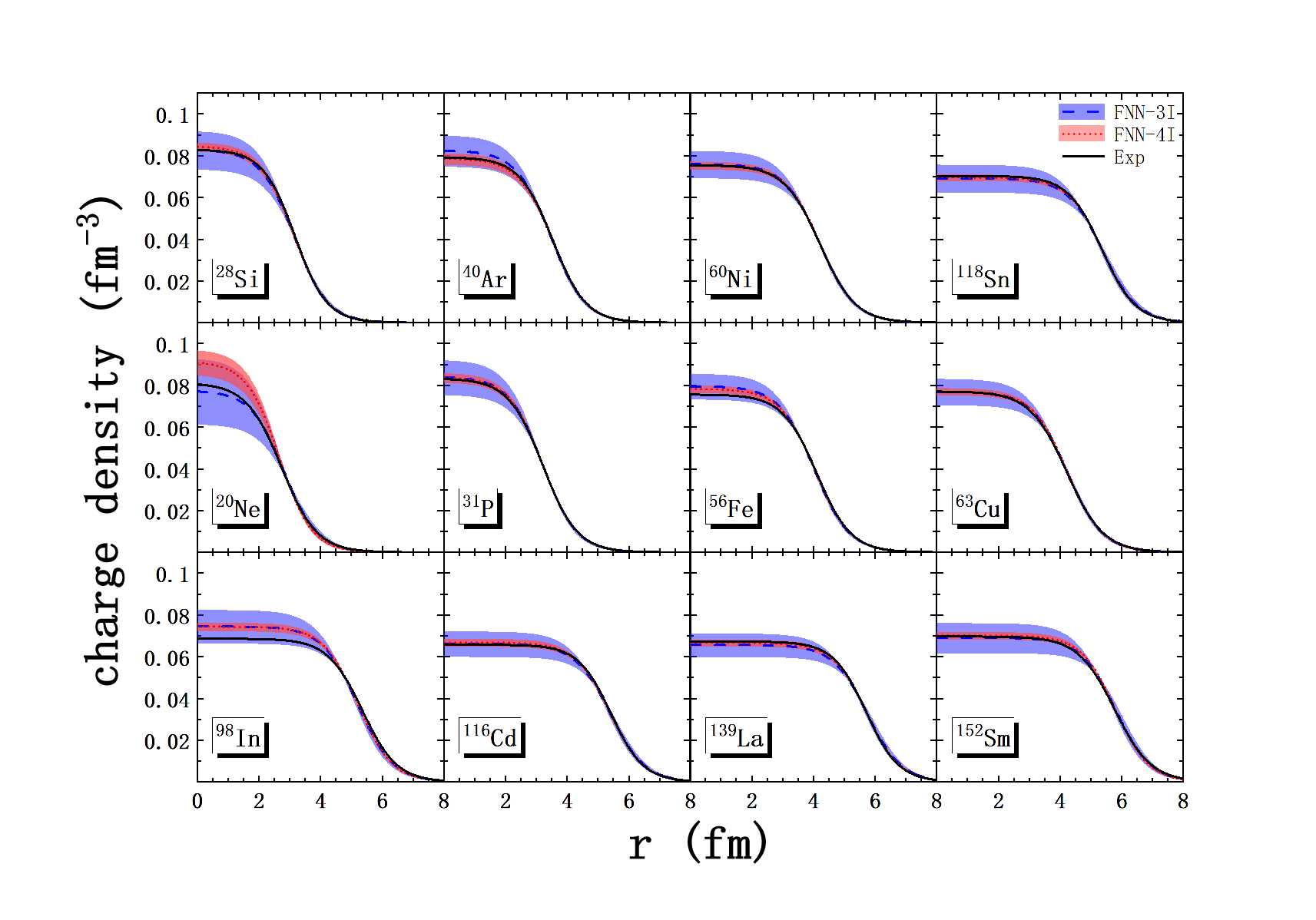}
\caption{(Color online) Charge density distributions obtained by FNN-I3 and FNN-I4.
The first row contains nuclei selected from the training set, and the others are nuclei selected from the validation set.
Density distribution determined by experimental parameters are shown by solid black lines.
Density distributions obtained by FNN-I3 and FNN-I4 methods are represented by blue hatched region and red hatched region, 
and their mean predicted values are marked by the blue dashed line and red dotted lines, respectively.}
\label{fig:Figure2}
\end{figure*}

The output of each layer of the neural network is marked as $[\textbf{\emph{a}}_1,\textbf{\emph{a}}_2, \textbf{\emph{a}}_3, 
..., \textbf{\emph{a}}_n]$, where $\textbf{\emph{a}}_1$ is the input of the network and $\textbf{\emph{a}}_n$ is the output of the network, and the number of 
each layer's neurons is labeled as $[N_1, N_2, N_3, ... , N_n]$. For the hidden layer, the outputs $\textbf{\emph{a}}_i$ are 
calculated by a formula:
\begin{equation}\label{eq3}
  \textbf{\emph{a}}_i = f(W_i \textbf{\emph{a}}_{i-1} +\textbf{\emph{b}}_i),
\end{equation}
where $W_i$ is the weight matrix between $i$-1th and $i$th layer with a shape of $N_{i} \times N_{i-1}$ and $\textbf{\emph{b}}_i$ 
is the bias vector of $i$th layer. The activation function $f$ carries out nonlinear mapping of the input, which is an important 
reason why FNN can fit most functions.In this paper, the activation functions of hidden layers are taken to be the hyperbolic tangent, tanh:
\begin{equation}\label{eq4}
  \tanh(x) = \frac{\sinh x}{\cosh x} = \frac{e^{x} - e^{-x}}{e^{x} + e^{-x}}.
\end{equation}

In the training procedure, we use the mean squared error (MSE) as the loss function:
\begin{equation}\label{eq5}
  Loss(\textbf{\emph{y}}_{tar}, \textbf{\emph{y}}_{pre}) = \frac{1}{N_s} \sum\limits^{N_s}\limits_{i=1}
  (\textbf{\emph{y}}_{tar}-\textbf{\emph{y}}_{pre})^2, 
\end{equation}
which is used to quantify the difference between model predictions $\textbf{\emph{y}}_{pre}$ and experimental values 
$\textbf{\emph{y}}_{tar}$. Here, $N_s$ is the size of the training set. The learning process is to minimize the loss function via 
a proper optimization method. Back-propagation algorithm with Levenberg-Marquardt \cite{LevenbergK.:1944, MarquardtD.W.:1963} for the training of the FNN was used.
FNN modifies its weights until an acceptable error level between predicted and desired outputs. We use stochastic gradient descent (SGD)
method \cite{RobbinsH.E.:2007} in this work to obtain the optimal parameters $W_i$ and $\textbf{\emph{b}}_i$ in the network. SGD is a popular 
alternative to gradient descent (GD), which is one of the most widely used training algorithms.

\begin{figure*}[!htb]  
  \includegraphics
    [width=0.9\hsize]
    {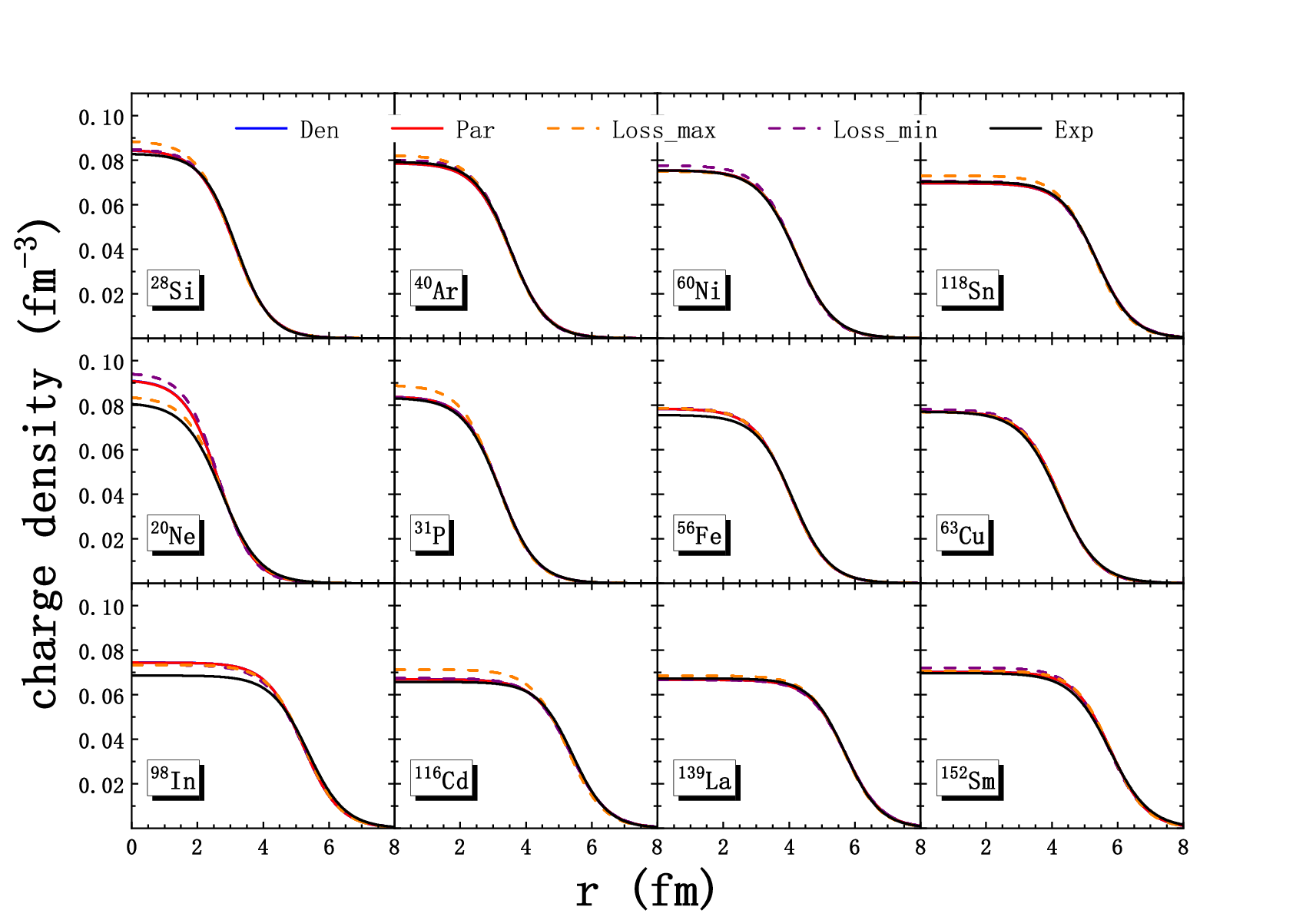}
  \caption{(Color online) The charge density distributions from different methods.
  The charge distributions obtained using experimental parameters are denoted by black solid lines. The distributions obtained 
  using the mean predicted density and the mean predicted parameter are represented by blue and red solid lines respectively. 
  And the distributions obtained by the parameters which minimize (maximize) the loss function are denoted by orange (purple) 
  dashed lines. The selection of nuclei is the same as in Fig. \ref{fig:Figure2}.}
  \label{fig:Figure3}
\end{figure*}

Since the charge radius of the nucleus can be described by $R = r_0 Z^{1/3}$ in most cases \cite{Chin-YuenT.:1957}, considering $Z^{1/3}$ 
in the input will help us to fit the parameter $c$ of 2pF. Sometimes, some experiments use an alternative radius formula \cite{HofstadterR.:1956,HofstadterR.andHahnB.andKnudsenA.W.andMcIntyreJ.A.:1954,YennieD.R.andRavenhallD.G.andWilsonR.N.:1954}: 
$R = r_0 A^{1/3}$, where $A$ is the mass number, so the $A^{1/3}$ has inputs apart from $Z, N, Z^{1/3}$ to ascertain the effect of this 
additional input. For simplicity, we use FNN-I3 and FNN-I4 to represent the FNN approaches with $\textbf{\emph{x}} = (Z, N, Z^{1/3})$ and 
$\textbf{\emph{x}} = (Z, N, Z^{1/3}, A^{1/3})$, respectively. In order to simultaneously predict two parameters $c$ and $z$ in 2pF, we adopt 
FNN with double hidden layer structure.

The experimental data of 2pF are taken from the Refs. \cite{DeJagerC.W.andDeVriesH.andDeVriesC.:1974,DeVriesH.andDeJagerC.W.andDeVriesC.:1987,FrickeG.andBernhardtC.andHeiligK.andSchallerL.A.andSchellenbergL.andSheraE.B.andDejagerC.W.:1995}. There are 86 nuclei left and their experimental 2pF data are taken in the 
dataset. If the 2pF parameter of a nucleus is obtained by multiple independent experiments, the most recent data will be adopted. 
\section{RESULTS AND DISCUSSION} \label{sec3}

In order to obtain reliable results, the neural network was trained repeatedly (1000 times), and each training will 
use randomly divided training set and validation set, among which the training set accounts for $80\%$ of the dataset (86 nuclei), and the 
validation set accounts for $20\%$. Moreover, at the beginning of each training, the network parameters will be randomly re-initialized. After each training, 
FNN-I3 and FNN-I4 will get the prediction results of parameters $c$ and $z$ on the validation set, which can be compared with the experimental 
results from Refs. \cite{DeJagerC.W.andDeVriesH.andDeVriesC.:1974,DeVriesH.andDeJagerC.W.andDeVriesC.:1987,FrickeG.andBernhardtC.andHeiligK.andSchallerL.A.andSchellenbergL.andSheraE.B.andDejagerC.W.:1995} to obtain a mean-squared-error (MSE).

Table \ref{tab:Table1} shows the MSE statistics of parameter $c$ and $z$ predictions of the FNN-I3 and FNN-I4 approaches for training set and 
validation set. As the result of FNN is affected by parameter initialization, some training results far deviate from the experimental value. In this paper, 
the upper limit of MSE is set for parameters: 0.2 for parameter $c$, and 0.005 for parameter $z$. It is worth mentioning that there are two nuclei ($^{98}\mathrm{In}$ and $^{102}\mathrm{Sb}$) in the 
dataset we used that are located far away from the others on the nuclide chart and are not stable nuclei. 
Removing these two nuclei from the dataset will not affect the results more than $\SI{0.002}{\femto\metre^2}$, and the relative error is less than $5\%$. Such errors can be considered as the result of removing the data from the dataset, and do not represent any particularities of the two nuclei. Results with MSE greater than the upper limit 
are eliminated. The number of predictions we accept is shown in the last column. The standard deviation (SD) is also shown in the table, which can be obtained by
\begin{equation}\label{eq6}
  SD = \sqrt{\frac{1}{N - 1} \sum_{i = 1}^N (y_i^{pre} - y^{tar})^2},
\end{equation}
where $N$ is the number of predictions adopted. $y_i^{pre}$ are the predicted values and $y^{tar}$ is the target value.

\begin{figure*}[!htb]  
  \includegraphics
    [width=0.9\hsize]
    {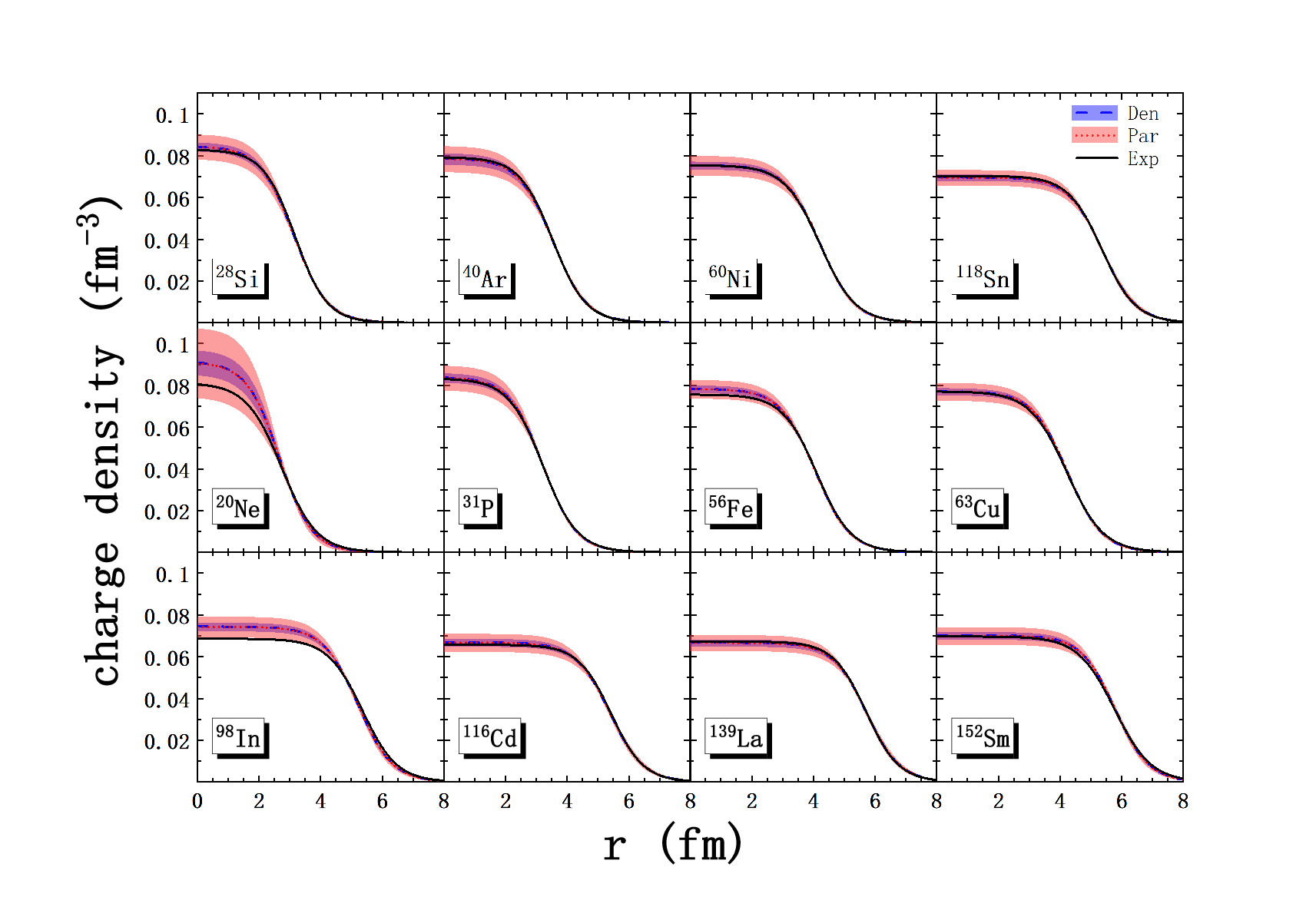}
  \caption{(Color online) The charge density distributions and error bands obtained by the average density distributions (Den) and the average parameters (Par) of multiple predicting, respectively,
  which are shown by blue dashed lines (blue hatched regions) and red dotted lines (red hatched regions).
  Density curves determined by experimental parameters are shown by solid black lines.
  The selection of nuclei is the same as in Fig. \ref{fig:Figure2}.}
  \label{fig:Figure4}
\end{figure*}

It can be clearly seen that for parameter $c$, the mean squared deviation of FNN-I3 is significantly larger than that of FNN-I4,  and the SD 
of FNN-I4 is smaller. Compared with FNN-I3, the mean MSE of FNN-I4 is reduced from $\SI{0.03913}{\femto\metre}^2$ to $\SI{0.02408}{\femto\metre}^2$ on training sets and from $\SI{0.03655}{\femto\metre}^2$ to 
$\SI{0.02286}{\femto\metre}^2$ on validation sets respectively, which is very precise for the predictions of parameter $c$. In fact, the nuclear charge radius is 
closely related to the mass numbers and proton numbers. Therefore, the above results indicate that the FNN approach is a reliable way to 
improve the accuracy of nuclear charge density distribution predictions based on experimental data by including known physics effects in the 
input layer. It should be noted that different combinations of physical quantities are tried as inputs, and it is found that the parameter $c$ 
is significantly correlated with $A^{1/3}$ and $Z^{1/3}$ (The Pearson correlation coefficients of $A^{1/3} - c$ and $Z^{1/3} - c$ are 0.9953 and 0.9969, respectively), 
while parameter $z$ was not sensitive to it (Pearson correlation coefficients are -0.1625 for $A^{1/3} - c$ and -0.1609 for $Z^{1/3} - c$, respectively). 
In view of this, the result of parameter $z$ is not considered in the following works.

To further evaluate the prediction ability of the FNN method, we show the MSE statistics of parameter $c$ in the case of fixed training set 
and validation set in Table \ref{tab:Table2}. The columns in Table \ref{tab:Table2} have the same meaning as those in Table \ref{tab:Table1}.
Since the value of parameter $c$ ranges from $\SI{2.4}{\femto\metre}$ to $\SI{7.0}{\femto\metre}$, it can be considered that the FNN approach is reliable.

Figure \ref{fig:Figure2} shows the prediction of the charge density distribution of some nuclei with fixed training set and validation set.
The average charge density distributions are obtained by applying the 100 FNN models with the best performance on the 
training set, which means the smallest loss function value, and apply it to the validation set to obtain the average charge density 
distributions. The error bands are obtained from the standard deviation of the density distribution values from the 100 FNN models.
It can be clearly seen from Fig. \ref{fig:Figure2} that the error bands of FNN-I4 are significantly narrower than those of FNN-I3, indicating 
that FNN-I4 has higher precision than FNN-I3. Moreover, in most cases, the average prediction distributions of FNN-I4 are closer to the 
density distribution obtained by experimental parameters than those of FNN-I3.In conclusion, FNN-I4 has higher accuracy than FNN-I3.

In order to illustrate the rationality of using the average density distribution as the prediction result in the above results, 
we compare the average result of multiple predictions and the result with the minimum (maximum) error on the validation set in Fig. \ref{fig:Figure3}. 
Since the 2pF model uses parameters to control the charge density distribution, in addition to averaging the charge density distribution 
curve to get the average predicted density distribution, it can also be obtained by averaging the predicted parameters. So Figure \ref{fig:Figure3} 
contains the prediction results obtained by these two averaging methods.

The following useful information can be obtained from Fig. \ref{fig:Figure3}: Firstly, the distribution of prediction density obtained by the two 
averaging methods almost coincides, which allows us to conveniently describe the prediction result without worrying about which method to use 
to obtain it; Secondly, the prediction results of the single minimum error network are not better than the results obtained by averaging. Since 
the effect of randomness is greatly reduced by averaging the results, the following works in this paper will use averaging method to obtain the 
prediction results. 

It is necessary to further explore the differences between the two averaging methods. Figure \ref{fig:Figure4} shows the average curves and error bands 
of charge density distributions obtained by the two methods on some nuclei, and the selection of these nuclei is the same as Fig. \ref{fig:Figure2}. 
It is obvious that the error bands obtained by averaging the density curves are narrower than those obtained by the other method, although the average 
curves of the two are very coincident. This is because even a small change in the parameter has a huge impact on the density distribution, so the 
uncertainty of the parameter will be amplified by mapping. Therefore, the predicted charge density distribution and error band will be obtained by 
means of average density curve later in this paper.

Since the interpolation ability of FNN-I4 has been verified, the entire sets are adopted as learning sets to assess the predictive power of the neural network. 
Because the second and fourth moments of the charge density distribution ($\left \langle r^2 \right \rangle$ and $\left \langle r^4 \right \rangle$) are important for understanding nuclear structure (for example, by means of statistical 
correlation analysis, it is demonstrated that the diffraction radius $R$ and surface thickness $\sigma$ are well determined by $\left \langle r^2 \right \rangle$ 
and $\left \langle r^4 \right \rangle$, especially for heavy nuclei \cite{ReinhardP.G.andNazarewiczW.andRuizR.F.G.:2020,NaitoT.andColoG.andLiangH.andRoca-MazaX.:2021}), the learning effect of FNN-I4 can be evaluated by $\left \langle r^2 \right \rangle$ and $\left \langle r^4 \right \rangle$.

\begin{figure}[!htb]  
  \includegraphics
    [width=0.9\hsize]
    {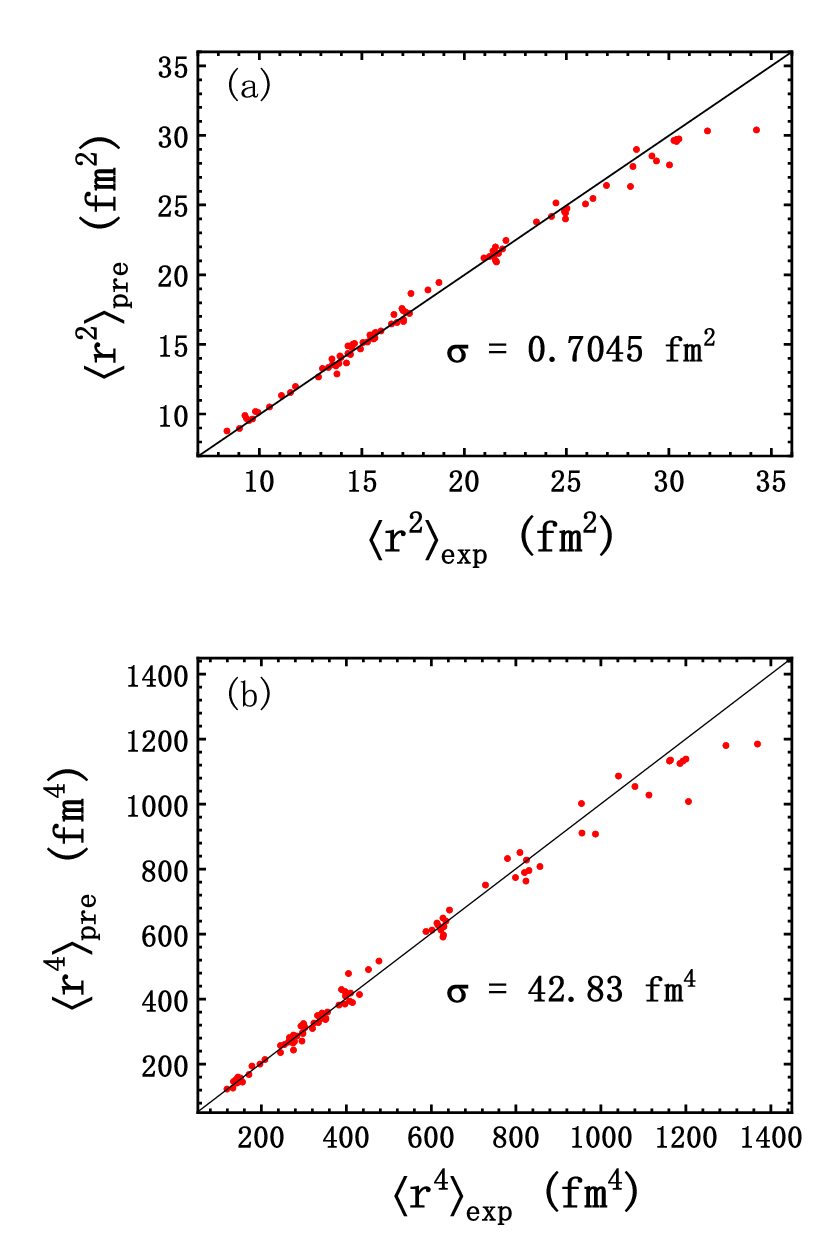}
  \caption{(Color online) The differences between the experimental second (a) and fourth (b) moments of charge density distribution and FNN training 
  results (red dot).}
  \label{fig:Figure5}
\end{figure}

The comparison of FNN-I4 predicted values and experimental values of $\left \langle r^2 \right \rangle$ and $\left \langle r^4 \right \rangle$ are shown in Fig. \ref{fig:Figure5}(a) and Fig. \ref{fig:Figure5}(b), 
respectively. Since the neural network directly outputs the parameters of the 2pF model, the predicted values of the second and fourth moments are actually calculated with high precision as follows\cite{ShabaevV.M.:1993}:
\begin{equation}\label{eq7}
  \left \langle r^2 \right \rangle = \frac{4\pi N_0c^5}{5}(1 + \frac{10}{3} \frac{\pi^2z^2}{c^2} + \frac{7}{3} \frac{\pi^4z^4}{c^4}), 
\end{equation}
\begin{equation}\label{eq8}
  \left \langle r^4 \right \rangle = \frac{4\pi N_0c^7}{7}(1 + 7 \frac{\pi^2z^2}{c^2} + \frac{49}{3} \frac{\pi^4z^4}{c^4} + \frac{31}{3} \frac{\pi^6z^6}{c^6}), 
\end{equation}
\begin{equation}\label{eq9}
  N_0 = \frac{3}{4\pi c^3} (1 + \frac{\pi^2z^2}{c^2})^{-1}. 
\end{equation}

As can be seen in Fig. \ref{fig:Figure5}, the FNN results agree exceptionally well with experimental values, especially for the light-medium mass nuclei. The rms 
deviations ($\sigma$) between the experimental charge radii and the results of FNN for $\left \langle r^2 \right \rangle$ and $\left \langle r^4 \right \rangle$ 
are $\SI{0.7045}{\femto\metre}^2$ and $\SI{42.83}{\femto\metre}^4$, respectively. Incidentally, the rms deviation of charge radii is $\SI{0.7041}{\femto\metre}$. 
The large error between the predicted results and the experimental results for heavier nuclei is acceptable because the data of heavier nuclei are few.

\begin{figure}[!htb]  
  \includegraphics
    [width=0.9\hsize]
    {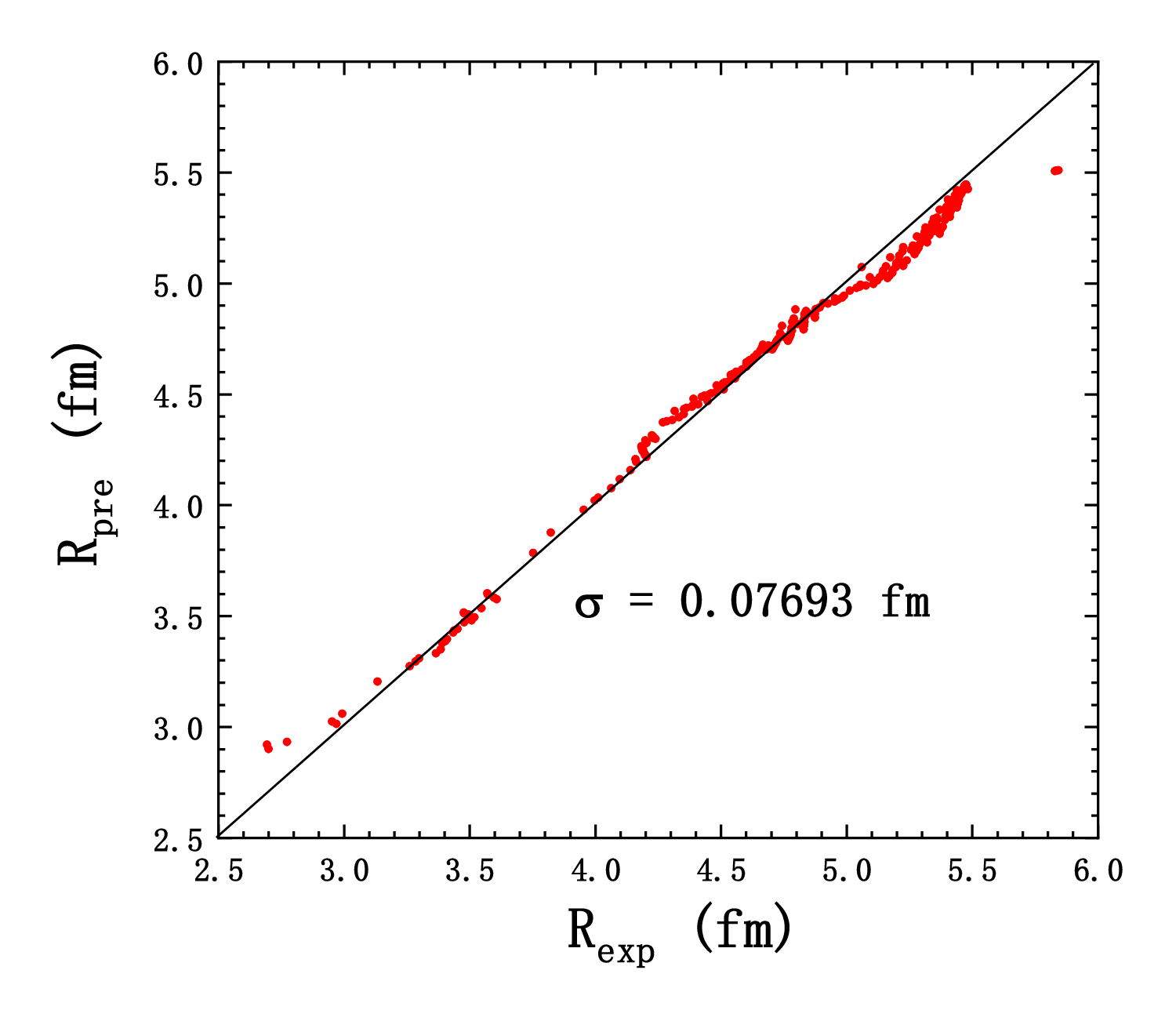}
  \caption{(Color online) Comparison of predicted rms charge radii($R_{\mathrm{pre}}$) and experimental values ($R_{\mathrm{exp}}$) for nuclei with available data.}
  \label{fig:Figure6}
\end{figure}

\begin{figure*}[!htb]  
  \includegraphics
    [width=0.7\hsize]
    {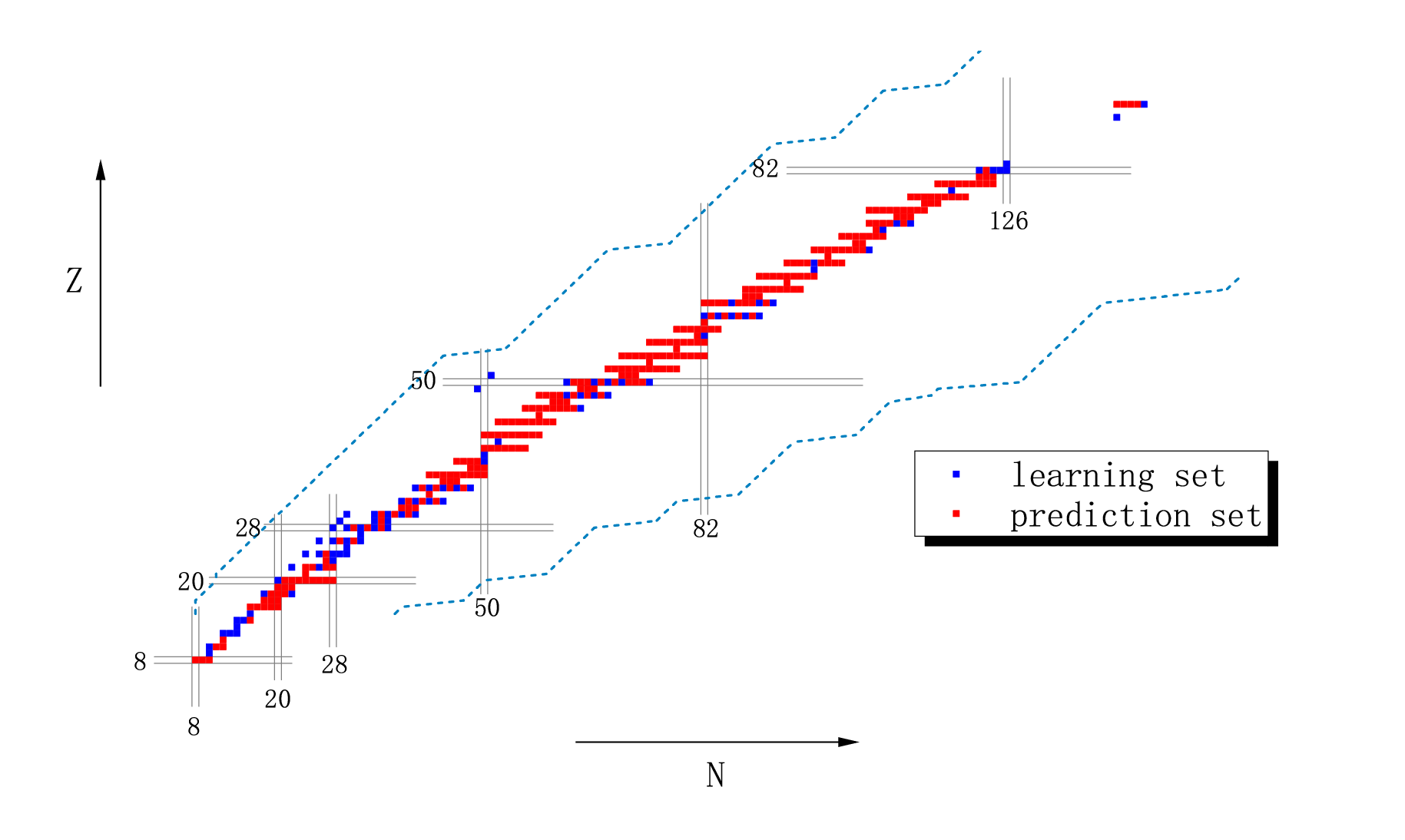}
  \caption{(Color online) learning set and prediction set in the nuclear chart.
  The learning set are shown by solid blue square, including all 86 nuclei, and the prediction sets are shown by red solid square with a total of 284.
  The two blue dashed lines are proton and neutron drip-lines, and the numbers represent magic numbers.}
  \label{fig:Figure7}
\end{figure*}

The distribution of the learning set and prediction set we used is shown in Fig. \ref{fig:Figure7}, where the solid blue square represents the learning set, 
with a total of 86 nuclei; The solid red squares represent the prediction set, which has 284 nuclei. The nuclei in the prediction set are selected as 
follows: taking the nuclei in the stable nuclei as the end points, this isotope chain is filled, and the filled nuclides are included in the prediction set, 
and of course, the nuclei that are included in the learning set are eliminated.
In Fig. \ref{fig:Figure6}, the charge radii of prediction set nuclei, obtained from the FNN-I4 approach, are compared with the experimental values. The data 
of experimental values are from Ref. \cite{AngeliI.andMarinovaK.P.:2013}. Among the 284 nuclei in the prediction set selected in this paper, a total of 230 nuclei have 
experimental charge radii.

\begin{figure*}[!htb]  
  \includegraphics
    [width=0.9\hsize]
    {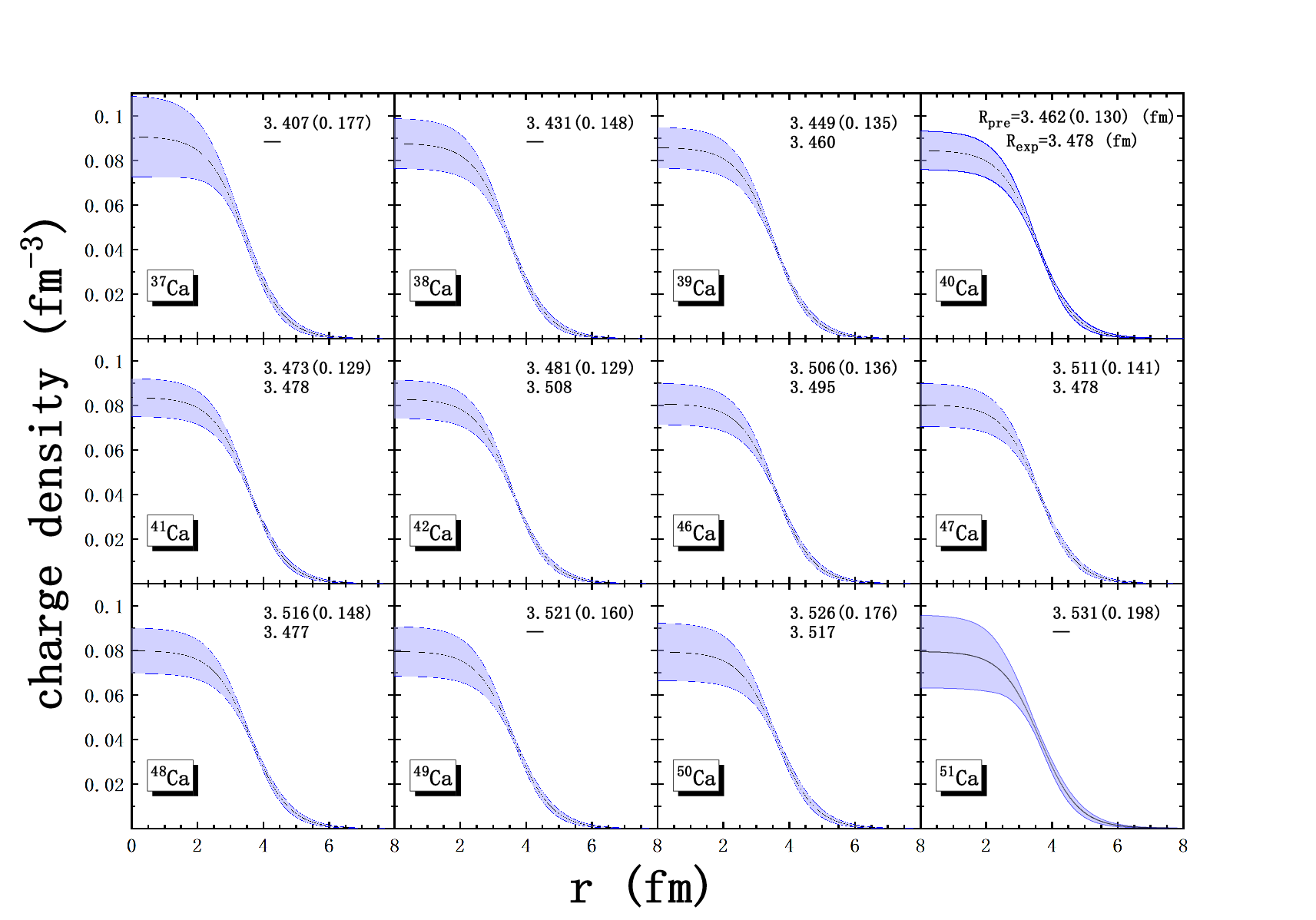}
  \caption{(Color online) Charge density distributions of $ ^{38-49}$Ca obtained by FNN-I4 method.
  The predicted charge radius ($R_{\mathrm{pre}}$) and experimental values ($R_{\mathrm{exp}}$) of each nucleus are also shown in the figure, 
  and the values in brackets after the $R_{\mathrm{pre}}$ are uncertainty.
  Those without experimental values are replaced by "—"}
  \label{fig:Figure8}
\end{figure*}

From Fig. \ref{fig:Figure6}, it is found that almost all points are near the line $y=x$, which means that the FNN-I4 predictions are very close to the experimental values, with the rms deviation 
($\sigma$) between the experimental charge radii and the results of FNN is $\SI{0.07693}{\femto\metre}$. Here are some nuclei that deviate far from the experimental values; they are either light 
nuclei near the oxygen isotope chain or heavy nuclei with $A > 208$. These nuclei may have different physical properties that are not well reproduced by the current 
learning set. With the increase of experimental data available for learning, the prediction accuracy of FNN-I4 in these regions may be improved. In summary, the 
FNN-I4 method has good prediction accuracy. The deviation of most predicted nuclei is less than $\SI{0.1}{\femto\metre}$, which indicates that the FNN-I4 method is a reliable method to predict nuclear charge distribution. 
Therefore, the interpolation ability of FNN-I4 method is also very promising.

We also tried to use the FNN-I4 method to predict the charge density distribution of calcium isotopes and get the charge radius from it. The experimental value of 
the charge radius was obtained from Ref. \cite{AngeliI.andMarinovaK.P.:2013}. Of the entire calcium isotope chain, 
only $^{40}\mathrm{Ca}$ was included in the learning set, with the other nuclei serving as the prediction set. Some of the results are shown in Fig. \ref{fig:Figure8}.
On the one hand, the uncertainty of the charge density in the nucleus center increases when it is far from the nucleus of the learning set, so only the prediction 
results near $^{40}\mathrm{Ca}$ can be regarded as reliable, which limits the extrapolation ability of the FNN-I4 method. On the other hand, most of the predicted 
values are in good agreement with the experimental values, except that it is difficult to predict the sudden drop of the charge radius around $^{48}\mathrm{Ca}$.
Considering that there is no corresponding situation in the existing learning set, add learning samples, which can help FNN approach extract physical information, 
is helpful and necessary to improve the prediction accuracy.
\section{summary}\label{sec4}

In summary, we have employed the feedforward neural network approach to predict nuclear charge density distribution, and it is clearly shown that the results 
obtained from FNN are in good agreement with the experimental data. Nuclear 2pF model parameters, taken from the Refs. \cite{DeJagerC.W.andDeVriesH.andDeVriesC.:1974,DeVriesH.andDeJagerC.W.andDeVriesC.:1987,FrickeG.andBernhardtC.andHeiligK.andSchallerL.A.andSchellenbergL.andSheraE.B.andDejagerC.W.:1995}, have been studied. 
By adding input variable of $A^{1/3}$ in the input layer, the machine learning method can accurately describe the 2pF model parameters of the nuclei (in the case of 
randomly dividing the dataset, the deviation in the training set is $\SI{0.02408}{\femto\metre^2}$ for parameter $c$ , and $\SI{0.00115}{\femto\metre^2}$ for parameter $z$), and has great results 
in the validation set ($\SI{0.02286}{\femto\metre^2}$ for parameter $c$ and $\SI{0.00188}{\femto\metre^2}$ for parameter $z$), which verifies the extrapolation ability of FNN. Then, without any 
experimental values, the charge density distribution (described by 2pF parameters) of 284 nuclei was calculated by FNN method. In addition, the density distribution of calcium isotopes 
and the corresponding charge radius were also obtained by FNN method.

So far, there are still a few experimental data on the ground state density distribution of spherical nuclei. Compared to traditional theoretical methods, the neural network 
method not only reduces the complexity of research, avoids complex multi-body problems, but also has great prediction ability with less computational cost. In the 
future, as the data available for neural networks to learn increases, the prediction ability will also be improved. Learning density distributions directly rather than model 
parameters may further help improve prediction.

\section{ACKNOWLEDGMENTS}\label{sec5}
This work is supported by the Natural Science Foundation of Jilin Province (Grant No. 20220101017JC), National Natural Science Foundation 
of China (Grant Nos. 11675063, 11875070 and 11935001), the Key Laboratory of Nuclear Data foundation (JCKY2020201C157), and the 
Anhui project (Z010118169).

\clearpage
\bibliographystyle{unsrt}
\bibliography{ref}
\end{document}